# Investigations of optical pumping for magnetometry using an auto-locking laser system

A. Pouliot[a], H.C. Beica[a], A. Carew[a], A. Vorozcovs[a], G. Carlse[a], B. Barrett[b] and A. Kumarakrishnan[a],
[a]Dept. of Physics and Astronomy, York University, 4700 Keele St., Toronto, ON, Canada M3J 1P3;
[b] L2PN, Labortoire Photonique, Numerique et Nanoscience, Institut d'Optique Graduate School, rue Francois Mitterrand, Talence 33400, France

## ABSTRACT

We have developed a versatile pulsed laser system for high precision magnetometry. The operating wavelength of the system can be configured to optically pump alkali vapors such as rubidium and cesium. The laser system consists of an auto-locked, interference filter stabilized, external cavity diode laser (ECDL), a tapered waveguide amplifier, and a pulsing module. The auto-locking controller can be used by an untrained operator to stabilize the laser frequency with respect to a library of atomic, molecular, and solid-state spectral markers. The ECDL output can be amplified from 20 mW to 2 W in continuous wave (CW) mode. The pulsing module, which includes an acousto-optic modulator (AOM), can generate pulses with durations of 20 ns and repetition rates of several MHz. Accordingly, the laser system is well suited for applications such as gravimetry, magnetometry, and differential-absorption lidar. In this work, we focus on magnetometric applications and demonstrate that the laser source is suitable for optically pumping rubidium vapor. We also describe numerical simulations of optical pumping relevant to the rubidium D1 and D2 transitions at 795 nm and 780 nm respectively. These studies are relevant to the design and construction of a new generation of portable, rubidium, spin-exchange relaxation-free (SERF) magnetometers, capable of sensitivities of 1 fT Hz$^{-1/2}$ [1].

## 1. INTRODUCTION

Alkali vapor cell magnetometers are used in wide ranging geophysical applications which involve detection of metal and mineral deposits. The high sensitivity of these magnetometers makes them suitable for wide-area, airborne surveys. In the SERF configuration these magnetometers potentially offer the highest sensitivity and operate without cryogenic cooling[1,2]. In recent work[3,4] we reviewed the characteristics of a versatile master-oscillator power amplifier (MOPA) system and preliminary studies of magnetometric applications. In this paper we describe the principles of SERF magnetometry in Sec. 2, we overview simulations of optical pumping of rubidium vapor on the D1 and D2 transitions in Sec. 3, and compare the results of these simulations with further experimental work on the D2 transition using the MOPA system in Sec. 4.

## 2. PRINCIPLES OF SPIN EXCHANGE RELAXATION-FREE MAGNETOMETRY

The fundamental principle of all alkali magnetometers is the measurement of the precession of the magnetic dipole moment of an atom in an external magnetic field. The frequency of this precession (i.e the Larmor frequency) is given by,

$$\omega_L = \frac{\vec{\mu_F} \cdot \vec{B}}{\hbar}, \quad (1)$$

where $\vec{\mu_F}$ is the atomic magnetic-dipole moment, $\vec{B}$ is the applied magnetic field and $\hbar$ is the reduced Planck constant. Magnetometers can be realized in either the time domain or frequency domain and we explain the operating principles of these devices using a simplified atomic level scheme. Figure 1 shows the level diagram for a rubidium magnetometer as it relates to the production and measurement of a magnetic field signal. A circularly-polarized pump laser on the D1 transition is used to optically pump rubidium vapor into a single magnetic ground state sublevel resulting in spin-polarization as shown in Fig. 1(a). After the sample is optically pumped into an extreme polarization projection (spin-polarized) the atoms will have a net magnetic dipole moment and will respond to external magnetic fields in a coherent manner. A magnetic field applied along the quantization axis defined by the pump laser would produce no time evolution of the initial magnetic sublevel populations since these sublevels are stationary states of the magnetic field Hamiltonian.

However, if the magnetic field is applied perpendicular to this quantization axis, the magnetic sublevel populations will oscillate at the Larmor frequency. This time evolution is analogous to the precession of a magnetic dipole moment about the axis of a magnetic field. The transmission of a linearly polarized probe laser through the optically pumped sample may be used to measure this oscillation. If the probe is off-resonance, the optically pumped sample will induce a Faraday rotation of the plane of polarization of the probe light corresponding to the spin polarization of the sample. For a resonant probe, the differential absorption of the orthogonal, circular components of the linearly-polarized probe (as shown in Fig. 1(b)) will allow the sample polarization to be measured.

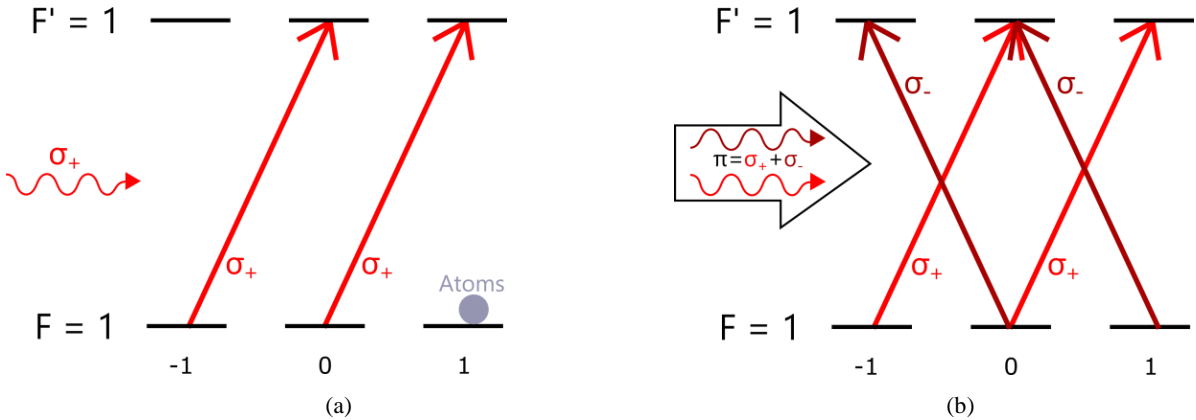

Figure 1. Simplified level diagram of a rubidium D1 transition. Demonstrating how the sample can be optically pumped into one $m_F$ state (a) and how the sample polarization can be measured using the differential absorption from a resonant, linear-polarized probe (b).

Figure 2 shows two practical implementations of a magnetometer. Figure 2(a) is well-suited to directly measuring the temporal Larmor oscillation[5], while Fig. 2(b) is suited to extracting the resonant frequency of the Larmor oscillation through a measurement of the lineshape of the magnetic dipole transition. In the time domain scheme, a short and intense pulse optically pumps the sample. The Larmor oscillations are recorded by measuring the differential absorption of a weak, linearly-polarized probe laser aligned at a small angle with respect to the pump.

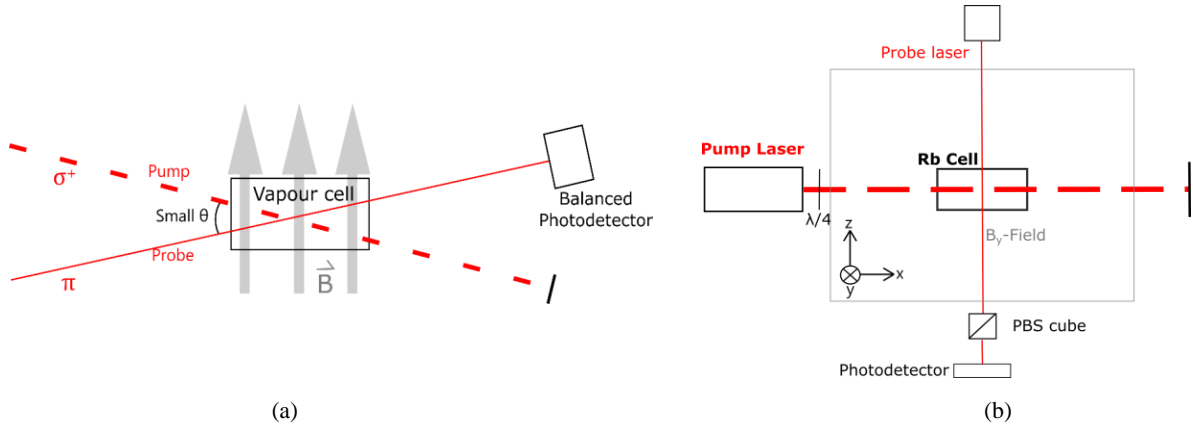

Figure 2: Schematic of a time-domain rubidium magnetometer (a) in which the pump and probe lasers are aligned at a small angle (the angle is exaggerated in the figure) and a frequency-domain magnetometer (b) in which the frequency of an amplitude-modulated pump laser is scanned to find the resonant Larmor frequency of the magnetic field being measured.

To detect the magnetic field in the frequency domain, the frequency of an AC magnetic field is scanned until it matches the resonance of the Zeeman shifted transition frequency due to the DC field being measured. The transmission through the sample allows the center frequency of the magnetic dipole transition to be determined[2]. A similar measurement can be made without AC field coils by scanning the frequency of an amplitude-modulated pump laser[1] as shown in Fig. 2(b). The signal-to-noise ratio of the line shape function can be further improved by amplitude modulating the probe laser and using lock-in detection at this modulation frequency[1].

The inherent sensitivity of any magnetometer is maximized by observing the Larmor oscillation for the longest possible duration. In practice, the decay time of the signal is limited first by the transit time of the atoms through the pumping volume. If the pumping volume is extended, the measurement time will ultimately be limited by the effect of wall collisions that decohere the Larmor oscillations. Therefore, the measurement time can be extended by using wall coatings that conserve angular momentum during collisional interactions with respect to any quantization axis. Under these conditions the measurement time is limited by radiation trapping which scrambles the atomic polarization. However, the addition of a small concentration of molecular (quenching) gas, such as $N_2$, with a broad range of resonant energies can ensure that collisional de-excitations dominate spontaneous emission and preserve the measurement time. In this regime, spin-exchange collisions between rubidium atoms, which result in transfer to atomic states that precess with the opposite phase, limit the time scale. Even so, this effect can be avoided by increasing the alkali density until the collisional frequency is large enough to re-initialize the phase of the Larmor precession resulting in the so-called spin-exchange relaxation-free (SERF) regime[1]. Since increasing the vapor density requires the cell to operate at a high temperature, it is necessary to use suitably heat-resistant wall coatings. The need for such coatings can be avoided by adding a high concentration of buffer gas. Collisions with the buffer gas will result in diffusive motion and effectively increase the transit time. The choice of buffer gas is dependent on its spin-destruction cross-section. Spin destruction collisions do not conserve angular momentum and impose the ultimate limit on the sensitivity of alkali magnetometers. Table 1 summarizes the collisional cross sections relevant to rubidium SERF magnetometers[6,7,8,9,10].

Table 1. Relevant cross sections of alkali magnetometry, all collisions refer to ground state atoms except where noted otherwise, quenching for Rb $P_{3/2}$ state is similar but not equal to $P_{1/2}$ cross sections.

| | |
|---|---|
| **Rb-Rb spin exchange cross section[6]** | $2 \times 10^{-14}$ cm$^2$ |
| **Rb-$N_2$ spin-destruction cross section[7]** | $1 \times 10^{-22}$ cm$^2$ |
| **Rb-He spin-destruction cross section[8]** | $9 \times 10^{-24}$ cm$^2$ |
| **Rb($P_{1/2}$)-$N_2$ quenching cross section[9]** | $6 \times 10^{-16}$ cm$^2$ |
| **Rb($P_{1/2}$)-He quenching cross section[10]** | $< 1 \times 10^{-22}$ cm$^2$ |

## 3. RATE EQUATION SIMULATIONS OF OPTICAL PUMPING

The simulation models the response of the atom to an applied optical field based on the semiclassical model of a multi-level atom[11]. An earlier rate equation description for a multilevel system is presented in Ref. 12. The rate of change of internal state populations can be approximated (up to second order in the applied field) by the set of rate equations

$$\frac{d\rho_{F,m_F}}{dt} = \sum_{H,m_H} \gamma_{H,F} \begin{pmatrix} H & 1 & F \\ m_H & q & m_F \end{pmatrix}^2 \rho_{H,m_H} + \Gamma \frac{|\chi_{H,F}|^2}{\left(\frac{\Gamma}{2}\right)^2 + \Delta_{H,F}^2} \begin{pmatrix} H & 1 & F \\ m_H & q & m_F \end{pmatrix}^2 (\rho_{H,m_H} - \rho_{F,m_F}), \quad (2a)$$

$$\frac{d\rho_{F',m_{F'}}}{dt} = -\Gamma \rho_{F',m_{F'}} - \Gamma \sum_{G,m_G} \frac{|\chi_{F',G}|^2}{(\Gamma/2)^2 + \Delta_{F',G}^2} \begin{pmatrix} F' & 1 & G \\ m_{F'} & q & m_G \end{pmatrix}^2 (\rho_{F',m_{F'}} - \rho_{G,m_G}). \quad (2b)$$

These rate equations describe the time evolution of the population $\rho$ in hyperfine ground state $|F, m_F\rangle$ (2a) and excited state $|F', m_{F'}\rangle$ (2b), in the presence of an applied optical field, where $F$ and $m_F$ are the total angular momentum and magnetic projection quantum numbers, respectively. These relations can be derived from the full set of density matrix equations[11,12], but since the goal of this simulation is to observe the steady-state atomic population distribution given by a particular driving field, all atomic coherences are ignored.

The second term in both equations describe the interactions of the atoms with the driving optical field, namely the stimulated absorption and emission processes characterized by the effective Rabi frequency $\chi_{H,G} = \langle H \| \boldsymbol{\mu} \cdot \boldsymbol{E} \| G \rangle / 2\hbar\sqrt{2H+1}$, and the detuning from atomic resonance $\Delta_{H,G} = \omega_{\text{las}} - \omega_{H,G}$. Here, $\boldsymbol{\mu}$ is the electric-dipole operator, $\boldsymbol{E}$ is the electric field strength, $\omega_{\text{las}}$ is the laser frequency and $\omega_{H,G}$ the atomic resonance frequency between hyperfine states $|H\rangle$ and $|G\rangle$. The Clebsch-Gordan coefficient $\begin{pmatrix} H & 1 & G \\ m_H & q & m_G \end{pmatrix}$ describes the transition strength between states with different angular momenta coupled by a photon with spin $q$, and is non-zero only for electric-dipole-allowed transitions. A time-dependent envelope function is used to model the amplitude of the pumping pulse over time.

The first term in Eqs. (2a) and (2b) describes relaxation due to spontaneous emission. The excited state decays with a total radiative rate $\Gamma$ (in vacuum), and the ground state is repopulated from all possible excited states with an effective rate proportional to the Clebsch-Gordan coefficient between initial and final states.

In the limit of a very large quenching gas concentration, such that relaxation occurs predominantly via collisions with the gas rather than spontaneous radiative decay, two effects are expected to alter the optical pumping process. These are the collisional broadening of a spectral line, and the decay channels opened by collisional decay which were not possible via electric-dipole transitions. A third effect, the shifting of the line center, is ignored here, since we limit ourselves to concentrations high enough where broadening makes this shift negligible. For simplicity, the collisional broadening is modeled by increasing the radiative rate (a more detailed model in which the radiative and collisional rates are treated separately will be presented elsewhere). The collisional transfer is modeled by replacing the first term in Eq. (2a) with terms of the form

$$\frac{d\rho_{F,m_F}}{dt} = \sum_{H,m_H} \frac{1}{\sum_G 2G+1} \Gamma \rho_{H,m_H} \tag{3}$$

where the denominator $2G+1$ gives the degeneracy of a given ground state manifold $|G\rangle$. This term allows all ground states to be populated by any excited state with equal probability, modeling the extreme case where radiative decays can be neglected.

Although this model is simplistic, it is effective at predicting the optical pumping of all magnetic, ground-state sublevels on long timescales at two pressure extremes, namely no quenching gas and a very large concentration of quenching gas (tens of atmospheres). A more detailed model of the collisional and radiative transfer gives more accurate predictions for all timescales and for intermediate concentrations of quenching gas. We also note that the model presented here does not include the optical coupling of the probe laser or the effect of the magnetic field.

We now model the optical pumping in the absence of quenching gases in both the D1 and D2 transitions of $^{87}$Rb at 795 nm and 780 nm respectively. The level diagrams and transition probabilities are shown in Fig. 3.

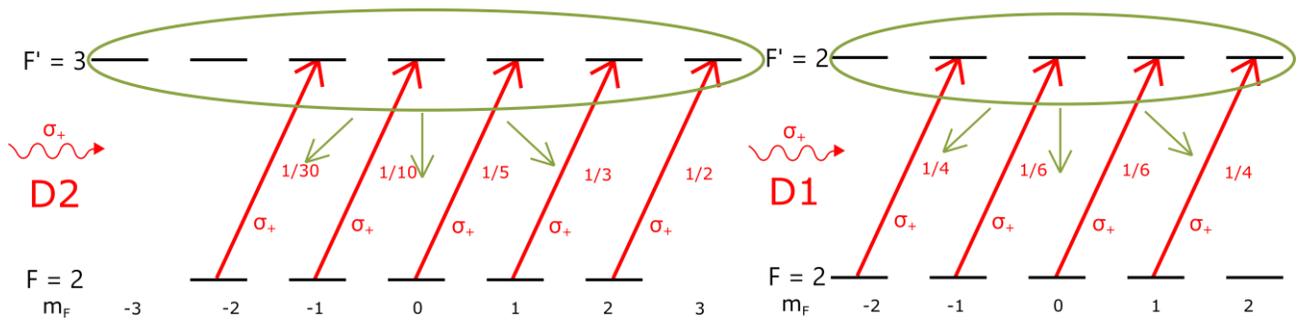

Figure 3. Optical pumping using $\sigma^+$ excitations. Relative strengths of each transition are shown in red, and green arrows represent examples of channels for collisional transfer. Without a quenching gas, optical pumping moves population towards the highest $m_F$ level. When quenching gas is added, all ground states are equally repopulated, so population collects in the state with the lowest probability of excitation, $m_F = -2$ in D2 and $m_F = 2$ in D1.

Figure 4(a) shows optical pumping in the D2 transition without buffer gases. Initially, the population is assumed to be equally distributed in all ground-state magnetic levels and a $\sigma^+$-polarized optical pumping field is applied on the F = 2 to F' = 3 transition. The results show that the populations of the F = 1 ground-states are unaffected, while the F = 2 populations are progressively pumped in to the extreme angular momentum state, $m_F$ = 2. Althought not shown in the figure, we confirm that the populations of the F' = 3 excited state are also pumped into the extreme ($m_F$ = 3) state. When the pumping field is turned off, the decay from this level results in a further increase of the $m_F$ = 2 ground state population.

Figure 4(b) shows optical pumping in the D1 transition without buffer gases. Here, the optical pumping field is resonant with the F = 2 to F' = 2 transition. Unlike in D2, the excited state can radiatively decay into the lower, F = 1, ground state which is ~6 GHz from resonance and is therefore no longer excited. As a result, the populations of the magnetic sublevels of the F = 1 state increase. The population remaining in the F = 2 ground-state is completely pumped into the $m_F$ = 2 sublevel. However, the population of this sublevel is lower than the populations of some of the F = 1 sublevels, and the overall polarization of the sample is significantly compromised. Over a very long pumping time, tails of the response curve would eventually pump atoms into $m_F$ = 2, however this would take a much longer pumping time (nearly 3 ms in this simulation). This state could be polarized quickly with the addition of a second pumping laser resonant with the F = 1 to F' = 1 transition. However, this would complicate the design of the magnetometer.

The results in Fig. 4 suggest the D2 transition would be much more appropriate for magnetometry without buffer gas. However, the dynamics of optical pumping change considerably with the addition of buffer and quenching gases and make the D1 transition the preferred choice, as we now show.

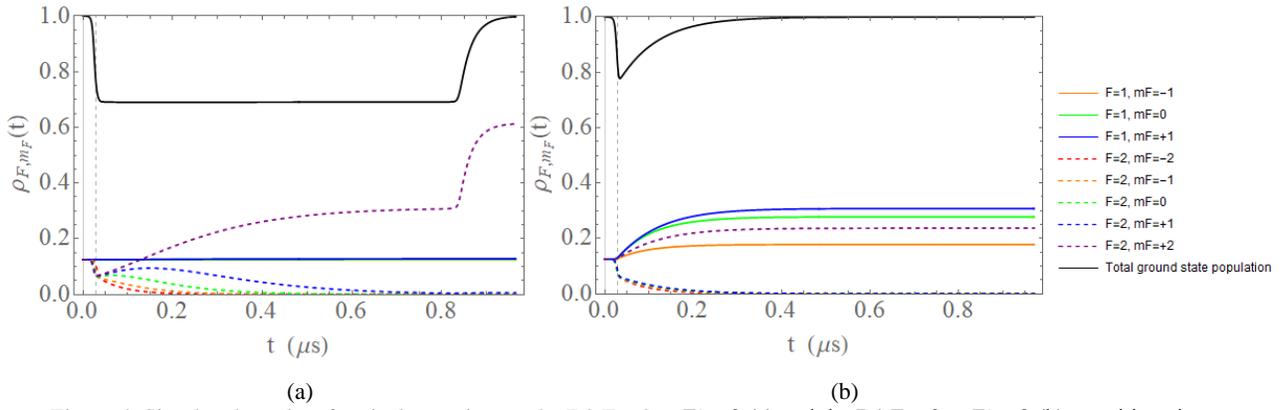

(a)      (b)

Figure 4. Simulated results of optical pumping on the D2 F = 2 to F' = 3 (a) and the D1 F = 2 to F' = 2 (b) transitions in [87]Rb using a $\sigma^+$-polarized pulse of duration 800 ns. These results were generated by numerically solving Eq. (2) assuming no buffer gas.

Figure 5(a) shows the results of optical pumping on the D2 transition in the presence of a relaxation rate of $3.8 \times 10^{12}$ Hz that models both the radiative and quenching processes. Under these conditions, there is sufficient collisional broadening to render the hyperfine ground and excited states degenerate. Electric-dipole selection rules continue to apply to all stimulated transitions. However, in the case of a very high quenching gas concentration, relaxation from each excited-state magnetic sublevel is mediated by collisions with the quenching gas and therefore populates all ground-state magnetic sublevels with equal probability. In contrast to Fig. 4(a), the $m_F$ = 2 sublevel is depopulated because it has the strongest transition probability to couple to the excited state with $\sigma^+$-excitation. The largest population collects in the $m_F$ = -2 sublevel which correspondingly has the weakest transition probability to the excited state via $\sigma^+$-excitation.

In contrast to Fig. 5(a), optical pumping on the D1 transition in the presence of quenching gas produces an overwhelming population accumulation in the $m_F$ = 2 magnetic sublevel, since this level is dark to $\sigma^+$-polarized light. Therefore, this system would be the ideal candidate for rubidium magnetometry.

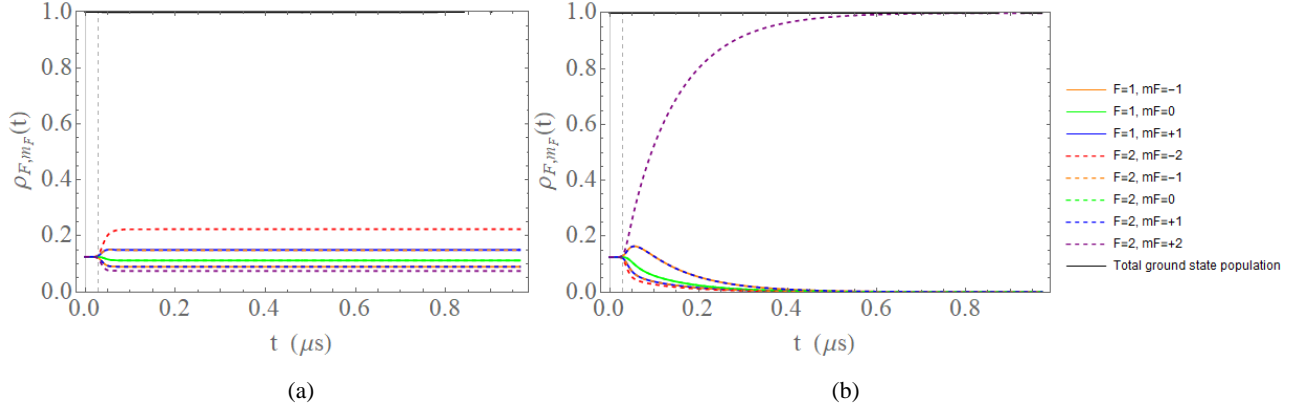

Figure 5. Simulated results of optical pumping on the D2 (a) and D1 (b) transitions in $^{87}$Rb with a decay rate of $3.8 \times 10^{12}$ Hz using a σ⁺-polarized pulse of duration 100 ns. As explained in the text, a more complete model of the collisional and radiative relaxations produces more accurate results.

## 4. EXPERIMENTAL RESULTS

The experimental setup is similar to the magnetometric configuration in Fig. 2 and in previous experiments[3,13]. An auto-locked external cavity diode laser is locked to the F = 2 to F' = 3 resonance in the D2 line (780 nm) of $^{87}$Rb using a saturated absorption spectrometer. The laser is split into a weak probe beam and an amplified optical pumping beam. The amplification is accomplished using a 2 W tapered amplifier waveguide (TA). Both pump and probe beams are pulsed independently using acousto-optic modulators (AOMs) with an operating frequency of 80 MHz. The pump and probe beams are overlapped through the rubidium vapor cell at a small angle of 10 milliradian. A quarter waveplate is used to circularly polarize the pumping beam. The probe beam is linearly polarized through the sample. A quarter waveplate and polarizing beam splitter after the sample are used to split the two components of the probe absorption onto a balanced photodetector. The detector consists of two Si PIN photodiodes with a 10 ns rise time. The current from the balanced detector is dropped across a resistor and amplified by two 20 dB RF amplifiers. A DC magnetic field is applied using two elliptical coils in a direction perpendicular to the propagation direction of the pump and probe beams. We now present results from two different rubidium cells, one containing room temperature rubidium vapor with natural abundances of $^{85}$Rb and $^{87}$Rb (length 5 cm) and another containing $^{87}$Rb vapor and 700 Torr of N$_2$ (length 8.5 cm) at a temperature of 40 °C. No specific efforts were made to shield the cells from the earth's magnetic field or ambient field gradients. The data were recorded using a 120 μs probe pulse and a 300 ns pump pulse with a repetition rate of 100 Hz.

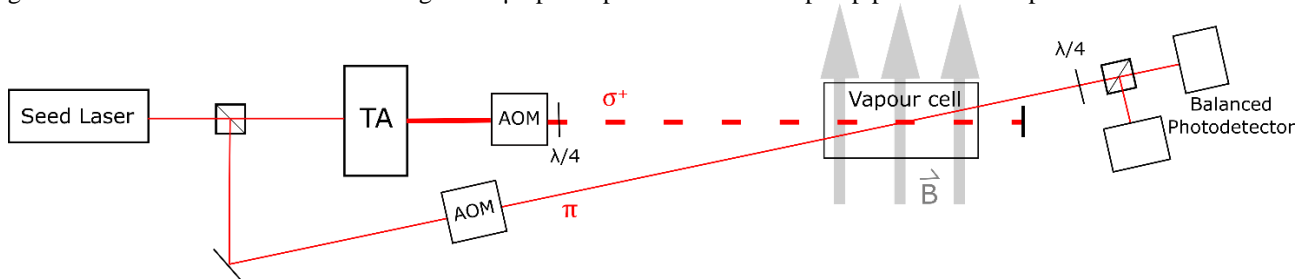

Figure 6. Experimental setup of the magnetometer. The peak power of the pump beam is 700 mW and the typical probe power is 1 mW.

Figure 7 shows a comparison of magnetometer signals from the two vapor cells in a magnetic field of approximately 3 G. The signal decay time in the buffer gas cell is approximately a factor of two larger, this should be able to be considerably extended by improving the signal to noise ratio of the buffer gas data. This behavior is consistent with the expectation that the transit-time-limited signal will be considerably extended due to diffusion through the buffer gas. Additionally, the rate of collisional de-excitation in the buffer gas cell exceeds the rate of spontaneous emission. Under these conditions radiation trapping is not expected to limit the decay time. The signal amplitude of the buffer gas data is notably larger due to the higher vapor pressure associated with high temperatures. Applying this technique to the vacuum cell begins to reduce decay time since it is limited by the transit-time of free-moving particles. Other factors that limit the decay time include the probe intensity, and the magnetic gradients across the cell length due to the field coils, the cell heaters, and magnetized materials near the cell. Based on the results of the simulations in Fig. 5, we expect that operating on the D1 transition under the same conditions will result in a larger signal amplitude, thereby permitting the signal decay to be observed over a longer timescale.

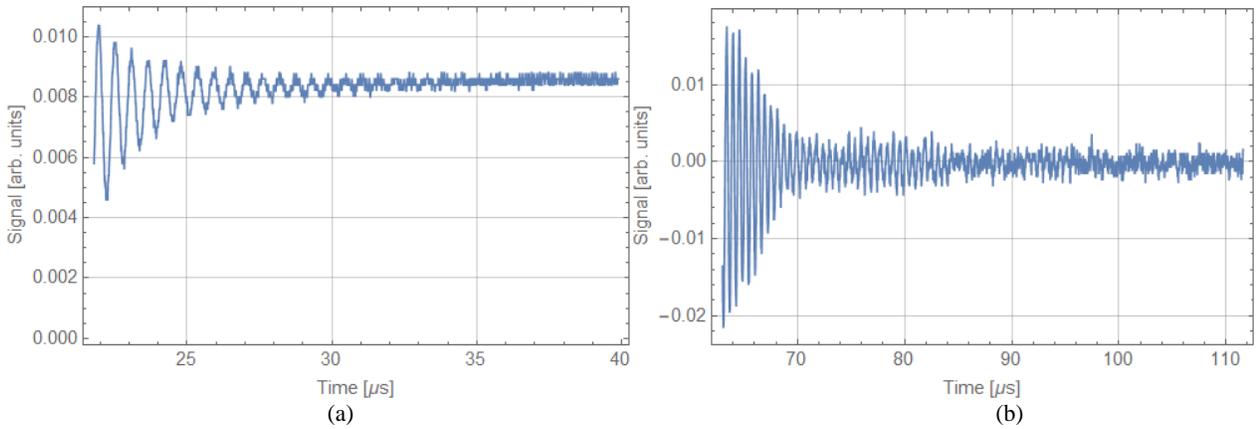

Figure 7. Time-domain magnetometric signals detected by measuring differential probe absorption in $^{87}$Rb. (a) Signal for a room temperature vapor cell with no buffer gas. (b) Signal for a cell with 700 Torr of $N_2$ at 40 °C. The applied magnetic field is approximately 3 G in both cases.

Figure 8 shows the Fourier transform of the signal in Fig. 7(b). The data are fit to a Lorentzian function added to a 1/f decay to account for noise. Datasets like this one show a single peaked spectrum but at lower magnetic fields (1 G) two peaks spaced by approximately 100 kHz are distinguishable. We currently attribute this effect to the presence of a magnetic field gradient across the cell.

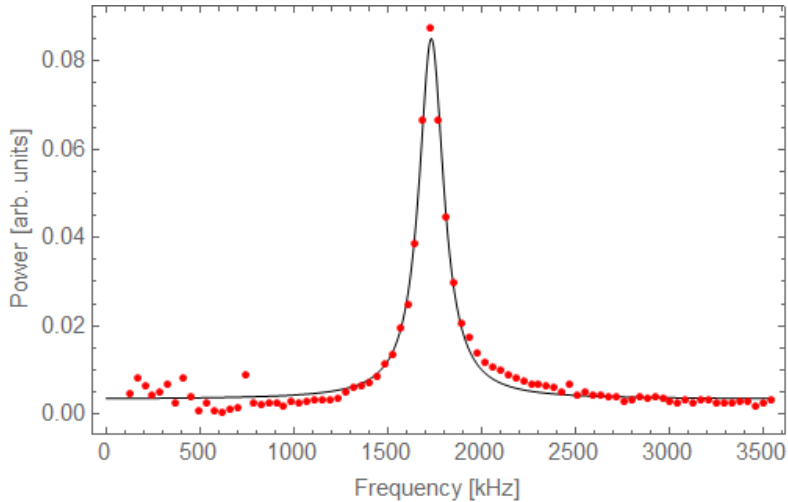

Figure 8. Fourier transform and fit corresponding to the time domain signal in Fig. 7(b).

Figure 9 shows the measured Larmor frequency as function of the applied magnetic field in the two vapor cells. In Fig. 9(a) a linear fit to the form $|sB_y + B_{y0}| + B_x$ gives a slope of $s = (716 \pm 5)$ kHz/G based on a current-to-field calibration of 42.3 G/A. This coil calibration was verified at the 10% level using a Hall sensor. The error bars in this fit have been weighted by the full-width at half maxima of the Lorentzian fits. The slope is comparable to the expected value of 700 kHz/G, and the measured deviation is entirely consistent with the uncertainty in the magnetic field due to the coil calibration and spatial gradients. Figure 9(b) shows similar data for the cell containing buffer gas. Here we find a slope of $(636 \pm 3)$ kHz/G. The uncertainty of each frequency measurement is reduced due to the extended timescale of the signal. However, the overall quality of the fit is comparable for both datasets. We attribute the deviation with respect to the expected slope of 700 kHz/G to the fact that the coils had to be displaced slightly to accommodate the heaters used for the buffer gas cell.

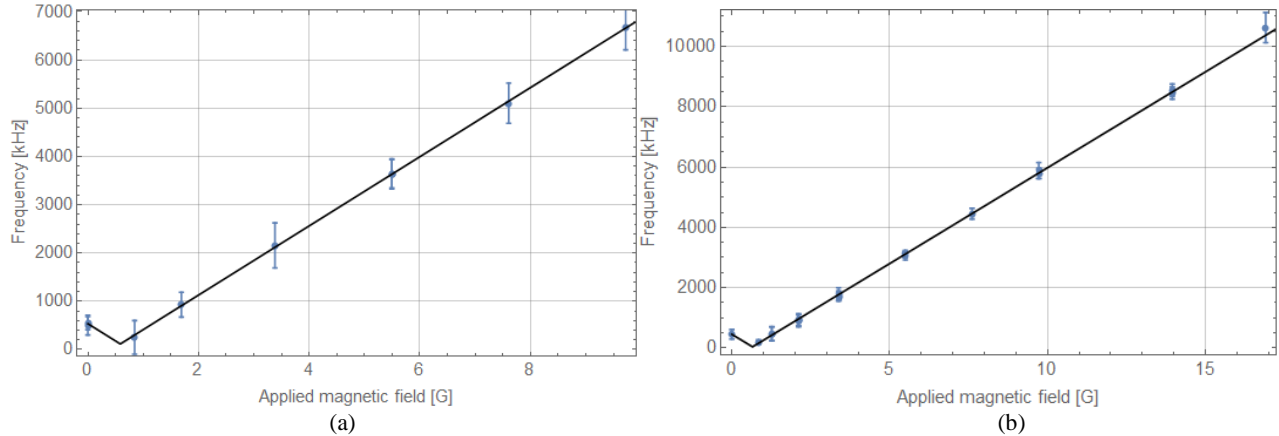

(a)                          (b)

Figure 9. Larmor frequency as a function of the applied magnetic field. (a) Data from a room temperature vapor cell with no buffer gas. (b) Data from a vapor cell containing 700 Torr of $N_2$ at a temperature of 40°C.

## 5. CONCLUSIONS

The preliminary data reported in this work demonstrates the suitability of the amplified auto-locked laser system for optically pumping rubidium vapor for magnetometric applications. We expect a significant improvement in the amplitude and duration of the signal by operating on the $^{87}$Rb D1 transition at 795 nm and by increasing the cell temperature to operate in the SERF regime. We also envision improving the optical pumping simulations using a density matrix approach that includes a more detailed model of the collision transfer, optical coherences, the effects of the probe absorption, and the magnetic field.